\pdfoutput=1
\documentclass[sigplan,nonacm]{acmart}

\AtBeginDocument{%
  }

\usepackage{amsmath}
\usepackage{mathtools}
\usepackage{algorithm}
\usepackage{algorithmic}
\usepackage{makecell}
\usepackage{multirow}
\usepackage{placeins}
\usepackage{tikz}
\usetikzlibrary{arrows.meta,positioning,calc}
\definecolor{planblue}{HTML}{245B91}
\definecolor{planorange}{HTML}{A95119}
\definecolor{plangreen}{HTML}{377153}
\definecolor{plangray}{HTML}{626A73}
\tikzset{
  planbox/.style={draw,rounded corners=2pt,line width=.65pt,align=center,
    inner sep=4pt,font=\small},
  planarrow/.style={-{Latex[length=1.7mm]},line width=.65pt},
}
\usepackage[capitalize,noabbrev]{cleveref}

\newcommand{\sysname}{Dyserve}

\setcopyright{none}
\renewcommand\footnotetextcopyrightpermission[1]{}

\begin{document}

\title{Serving Agentic Workflows with a Physical-Plan Compiler and Adaptive Runtime}
\author{Jiayi Qian}
\authornote{Corresponding authors: Jiayi Qian
(\texttt{jiayiqian@gatech.edu}) and Zishen Wan
(\texttt{zishen.wan@columbia.edu}).}
\affiliation{%
  \institution{Georgia Institute of Technology}
  \city{Atlanta}
  \state{GA}
  \country{USA}}

\author{Yichong Zhang}
\affiliation{%
  \institution{Harvard University}
  \city{Cambridge}
  \state{MA}
  \country{USA}}

\author{Hanchen Yang}
\affiliation{%
  \institution{Georgia Institute of Technology}
  \city{Atlanta}
  \state{GA}
  \country{USA}}

\author{Chun Tao}
\affiliation{%
  \institution{Intel}
  \city{Santa Clara}
  \state{CA}
  \country{USA}}

\author{Souvik Kundu}
\affiliation{%
  \institution{Intel}
  \city{Los Angeles}
  \state{CA}
  \country{USA}}

\author{Zishen Wan}
\authornotemark[1]
\affiliation{%
  \institution{Columbia University}
  \city{New York}
  \state{NY}
  \country{USA}}

\author{Tushar Krishna}
\affiliation{%
  \institution{Georgia Institute of Technology}
  \city{Atlanta}
  \state{GA}
  \country{USA}}
\renewcommand{\shortauthors}{Qian et al.}

\begin{abstract}

Efficient serving of agentic workflows requires selecting each LLM node's model, verification policy, and backend to balance output quality, latency, and throughput. These assignments must also adapt to changes in serving load. Existing approaches address parts of this problem through model routing, verifier placement, and backend scheduling. However, independent optimization overlooks their dependencies: model and backend choices determine verification cost, while verification changes the quality-cost trade-off among models. Ignoring these interactions
can waste serving resources and degrade workflow performance.

To address this problem, we propose \textbf{\sysname{}}, which provides the missing workflow physical-planning layer between orchestration and
model serving through compiler-runtime co-design. Our design is guided by three observations: planning headroom is request-dependent; node vulnerability, the impact of local errors on final correctness, depends
on position and task type; and serving load changes the cost of a plan during execution. For each request, its profile-guided compiler jointly
selects node implementations and prepares pressure-specialized variants
for the materialized workflow. The runtime selects variants using live
backend pressure and updates only undispatched assignments, without
invoking the optimizer on the load-change path.

Across four agentic workloads, \sysname{} improves accuracy by
\textbf{3-9} percentage points with \textbf{1.1-6.8}$\times$ mean-latency speedups
over the highest-accuracy evaluated baseline for each workload.
On a burst trace, variant switching raises the fraction of correct,
on-time completions from \textbf{18.1\%} to \textbf{67.2\%} relative to admission-only
execution.

\end{abstract}

\maketitle
\pagestyle{plain}
\thispagestyle{plain}

\section{Introduction}
\label{sec:intro}

Large language models are increasingly used for code generation,
scientific reasoning, and tool-augmented problem solving. Many such tasks
cannot be completed reliably by a single model invocation.
Agentic applications~\cite{yao2023react,wu2024autogen,hong2024metagpt}
organize LLM and tool calls into workflows that retrieve information,
validate intermediate results, and revise earlier outputs. This shift
changes the unit of serving from an isolated call to an end-to-end
computation whose calls play different roles in producing the final result.

We focus on workflows materialized before execution as directed acyclic
graphs~\cite{hong2024metagpt,zhang2025multiagentarchitecturesearch,
wang-etal-2025-evoagentx,zhang2025aflow}.
The graph fixes operations and dependencies, but not each LLM node's
physical implementation. On comparable hardware, smaller models can
reduce latency and memory use, potentially supporting larger batches
and higher throughput, but may reduce output quality. Verification
policies use additional model calls to check or refine a node's output,
potentially improving reliability at additional serving cost.
The workflow-wide assignment of models, verification policies, and
backends forms a \emph{workflow physical plan}. Its choices jointly
affect final correctness, critical-path latency, and fleet demand.

\begin{figure}[t]
  \centering
  \includegraphics[width=\linewidth]{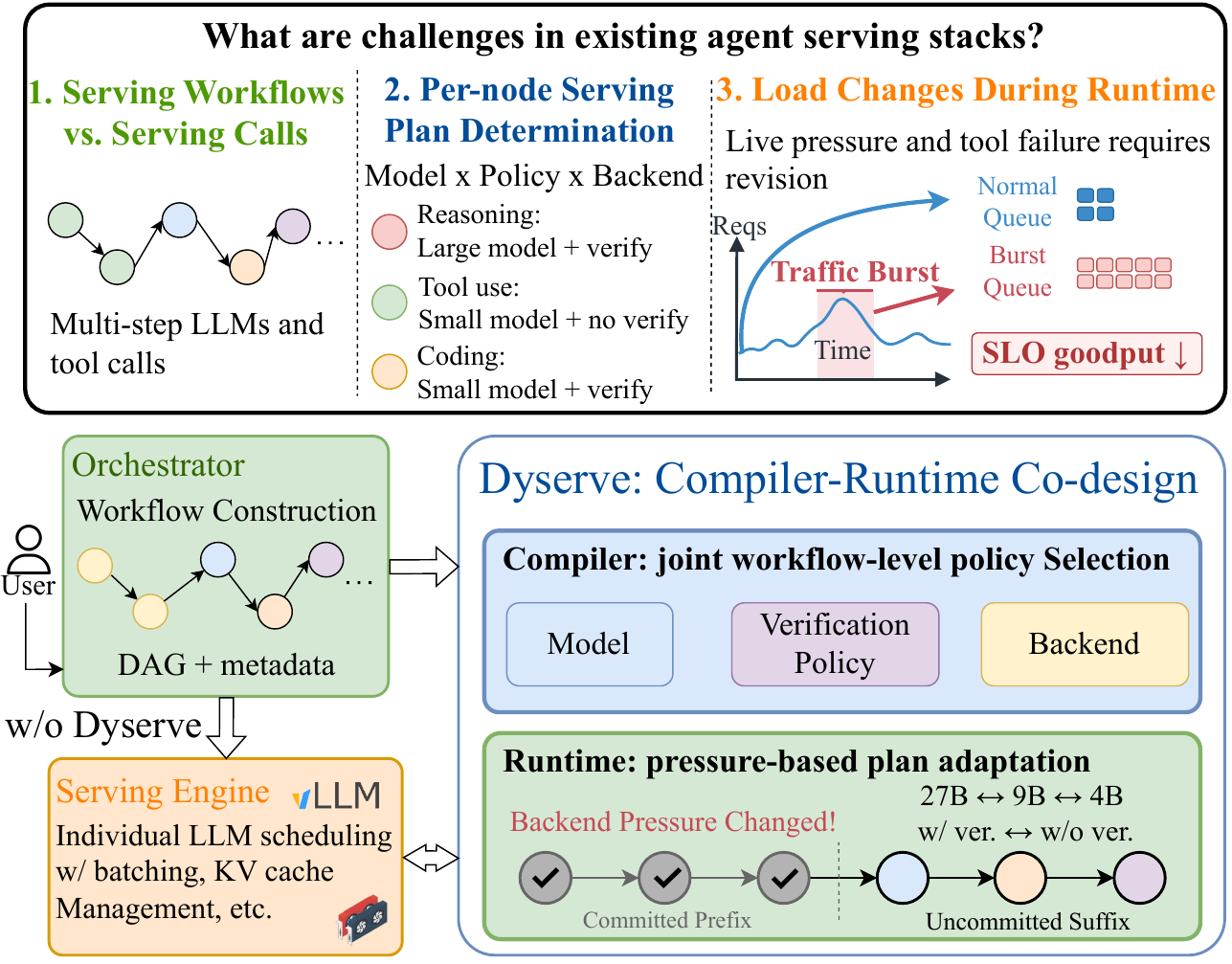}
  \caption{\sysname{} bridges workflow orchestration and model serving
  through joint physical-plan compilation and runtime adaptation.}
  \label{fig:stack}
  \vspace{-5pt}
\end{figure}

The conventional stack splits responsibility for this physical plan
between orchestration and serving (\cref{fig:stack}). The orchestrator
knows the workflow and application state, but lacks direct visibility
into backend queues and model-specific service conditions. Serving
engines manage batching, cache reuse, and scheduling, but typically receive calls after their model and verification choices have been fixed.
Model routers~\cite{ong2025routellm,chen2024frugalgpt}, application-aware
schedulers~\cite{lin2024parrotefficientservingllmbased,luo2026agentix,shahout2026orla},
and verifier placers~\cite{ro2025sherlock} address different parts of this
problem. The missing responsibility is to evaluate their combined workflow-level effects and
maintain the resulting assignment as one revisable execution plan.


However, through our profiling we found that simply composing a model router, a verifier placer, and a backend
scheduler as independent control loops does not resolve this ownership
gap. Our characterization reveals request-dependent quality-latency
headroom, but also shows that \textbf{physical-plan choices are coupled
both within nodes and across the workflow}. The selected model and
backend determine verification's cost, while verification changes the
quality-cost trade-off among model choices. Shortening one parallel
branch can also make verification on another branch delay the workflow.
These interactions require complete implementations to be evaluated
under a common workflow-level objective (\cref{sec:motivation}).

The value of improving a node also depends on the consequence of its
errors. Our fault-injection study shows that \textbf{node vulnerability
depends on both structural position and task type}. Vulnerability measures
the loss in final correctness caused by an injected node fault.
Nodes in the same structural group exhibit different vulnerability across
task types. We therefore extend topology-based vulnerability
analysis~\cite{ro2025sherlock} with task-type conditioning to guide model
and verification allocation (\cref{sec:motivation:coupling}).

Even a jointly constructed physical plan can become unsuitable during
execution. Our serving experiments show that \textbf{a fixed plan's
latency can change substantially across workload mixes}.
Moving an undispatched node to a smaller model can reduce delay under
congestion, but also changes the cost and value of verification. Tool
failures can require request-specific repair. Adaptation must therefore
coordinate the remaining assignments with low planning overhead while
preserving dispatched work (\cref{sec:motivation:lifecycle}).

\begin{figure}[t]
  \centering
  \includegraphics[width=\linewidth]{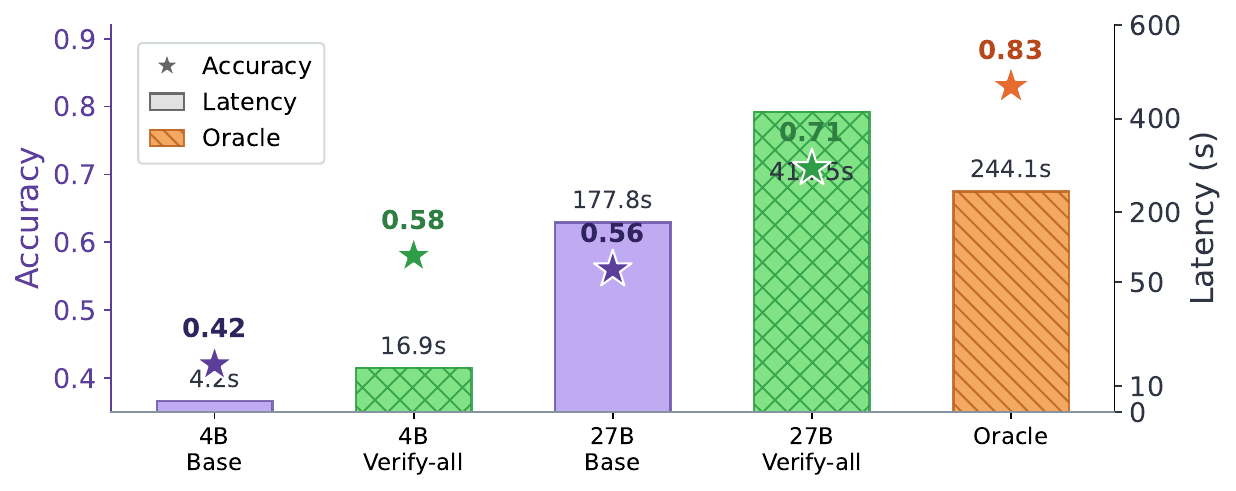}
  \caption{The evaluated plans expose request-dependent quality-latency headroom on LiveCodeBench.}
  \label{fig:physical-plan-space}
  \vspace{-5pt}
\end{figure}

We therefore propose \textbf{\sysname{}}, a workflow
physical-planning layer between the orchestrator and a heterogeneous
model-serving fleet (\cref{sec:layer}). Its compiler-runtime co-design
separates per-request plan construction from load-driven plan selection.
The compiler constructs coordinated alternatives using the complete
workflow and reusable profiles; the runtime selects among them using
live backend pressure. This division maintains workflow-wide coordination
without invoking the optimizer on the load-change path.

At admission, the compiler solves a workflow-wide integer linear program
(ILP) per request. It combines model- and backend-dependent
verification costs, workflow precedence, and our position- and
task-type-conditioned vulnerability estimates to select node
implementations and prepare pressure-specialized variants
(\cref{sec:admission}). During execution, the runtime installs a selected
variant at a node-boundary safe point, updating only undispatched
assignments. Completed and in-flight invocations retain their bindings.
Successful bounded tool-failure recovery can trigger one suffix
recompilation (\cref{sec:runtime}). The orchestrator retains application
control flow, and serving engines retain backend-local execution.

We implement \sysname{} over 4B, 9B, and 27B model backends and evaluate
it on LiveCodeBench~\cite{jain2024livecodebench},
GAIA~\cite{mialon2024gaia},
ComplexFuncBench~\cite{zhong2025complexfuncbench}, and
SWE-bench~\cite{jimenez2024swebench}.
Across these workloads, \sysname{} improves accuracy by \textbf{3-9} percentage
points and achieves \textbf{1.1-6.8}$\times$ mean-latency speedups over each
workload's highest-accuracy evaluated baseline. On a burst trace,
runtime variant selection raises SLO goodput, the fraction of correct,
on-time completions, from \textbf{18.1\%} to \textbf{67.2\%} relative to admission-only
execution (\cref{sec:eval}).

\begin{figure}[t]
  \centering
  \includegraphics[width=\linewidth]{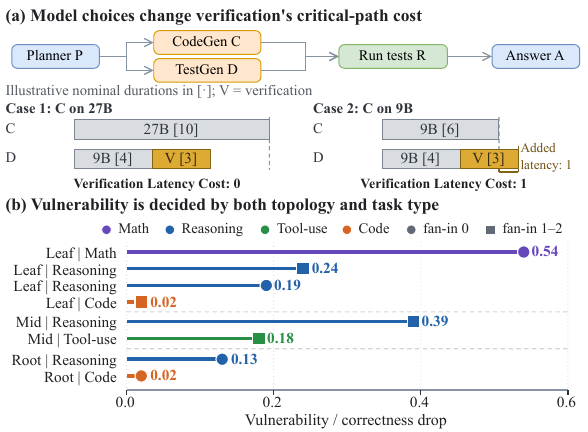}
  \caption{Workflow-dependent latency and error impact.
  (a) $C$'s model affects the same verification's added latency.
  (b) Vulnerability varies by task type within a structural group.}
  \label{fig:physical-plan-dependencies}
  \vspace{-5pt}
\end{figure}

Our contributions are:

\begin{itemize}

\item We characterize request-dependent physical-plan headroom,
cross-node decision dependencies, task-type-dependent vulnerability,
and sensitivity to serving conditions. These observations motivate a
physical-planning layer that owns the complete plan throughout execution
(\cref{sec:motivation,sec:layer}).

\item We develop a profile-guided compiler with position- and
task-type-conditioned vulnerability analysis. It jointly selects model,
verification, and backend assignments and generates pressure-specialized
plan variants (\cref{sec:admission}).

\item We build an adaptive runtime that switches variants at
node-boundary safe points while preserving dispatched bindings, and
recompiles the suffix after successful bounded tool-failure recovery
(\cref{sec:runtime}).

\item We evaluate \sysname{} on four agentic workloads, demonstrating
  \textbf{3-9} percentage-point accuracy gains and \textbf{1.1-6.8}$\times$
  mean-latency speedups over each workload's highest-accuracy evaluated
  baseline. On a burst trace, variant switching improves SLO goodput
  from \textbf{18.1\%} to \textbf{67.2\%} versus admission-only execution
  (\cref{sec:eval}).
\end{itemize}

\section{The Physical-Planning Gap}
\label{sec:motivation}

To motivate a dedicated workflow physical-planning layer, we ask
three questions. Does the physical-plan space provide substantial
and request-dependent quality-latency headroom? Can independent
model, verification, and backend decisions exploit this headroom?
How should the plan respond when serving conditions change during
execution? We use these questions to derive the requirements for
\sysname{}.

\subsection{Physical-Plan Headroom Is Large and Request-Dependent}
\label{sec:motivation:headroom}

A logical workflow fixes its operators and dependencies but leaves
open each LLM node's model, verification policy, and serving backend.
These alternatives create a large physical-plan space with room
to improve on a uniform configuration.

We therefore evaluate a bounded set of candidate plans offline and
construct a measured per-request oracle. For each request, the
oracle selects in hindsight the best measured outcome among the
candidate plans executed for that request.
\Cref{fig:physical-plan-space} compares this oracle with the
evaluated fixed configurations on LiveCodeBench. The fixed
configurations span from 0.42 accuracy at 4.2 seconds to 0.71
accuracy at 414.5 seconds. The oracle reaches 0.83 accuracy at
244.1 seconds, improving both accuracy and mean latency relative
to the highest-accuracy fixed configuration. The offline search
therefore exposes a quality-latency point that no single
evaluated fixed configuration attains.

The evaluated plans are complementary across requests: a plan that
succeeds on one request may fail or incur unnecessary cost on another.
The oracle is an offline upper bound obtained by evaluating multiple
candidate plans for each request. Because the full per-node plan space
grows combinatorially, such outcome-based plan evaluation is infeasible
on the serving path.

\noindent\textbf{Observation 1: Physical-plan headroom is substantial
and request-dependent.}
The evaluated candidates expose useful quality-latency headroom,
and the preferred plan depends on the request. Exploiting this
opportunity requires efficient per-request plan selection from
information available at admission.

\subsection{Physical-Plan Decisions Must Be Coordinated}
\label{sec:motivation:coupling}

Efficient per-request selection must account for dependencies both
within a node's implementation and across the workflow. At one node,
a verification policy introduces checking or repair calls whose cost
depends on the selected model and backend. Verification also changes
the quality-cost trade-off among models. A router, verifier placer,
and backend scheduler operating as independent control loops can
therefore miss useful combinations. The complete model-verifier-backend
implementation is the unit of evaluation.

\noindent\textbf{Verification's critical-path cost depends on other nodes.}
Even complete implementations cannot be evaluated only in isolation.
\Cref{fig:physical-plan-dependencies}(a) illustrates parallel code
generation at $C$ and test generation at $D$. Changing $C$ from a
27B to a 9B model shortens its nominal duration from 10 to 6 units;
$D$ retains the same 9B execution of 4 units and verification of 3
units. Comparing execution with and without verification within each
case, verification adds $\max(10,7)-\max(10,4)=0$ units in Case~1,
but $\max(6,7)-\max(6,4)=1$ unit in Case~2. These are illustrative
nominal times, not measurements; the common planner, test-execution,
and answer stages are excluded from both fork-join comparisons.
Thus, a model choice at one node changes the marginal latency cost
of verification at another. Verification consumes additional serving
resources in both cases, including when it fits within branch slack.
Workflow-level planning must distinguish this serving work from the
added critical-path latency; \cref{sec:walkthrough} illustrates how
these costs affect the joint choice.

\noindent\textbf{Vulnerability depends on both position and task type.}
Sherlock~\cite{ro2025sherlock} measures node vulnerability through
counterfactual fault injection and derives a verifier-placement policy
based on structural position and fan-in. It also identifies node
functionality as a possible extension. We develop a joint vulnerability
profile over position and node task type, represented by the required-skill
tag. This distinguishes the consequences of errors at structurally
similar nodes that perform different tasks, rather than assigning
protection priorities from position alone.

We compare paired unperturbed and fault-injected executions, grouped
by structural role, required skill, and fan-in. Among pairs with a
correct unperturbed result, vulnerability is the fraction whose final
result becomes incorrect after injection. It measures the consequence
of the injected fault, not the probability that the node naturally
produces an error. In \cref{fig:physical-plan-dependencies}(b), color
denotes task type and marker shape denotes fan-in. Intermediate reasoning
nodes with fan-in one or two have vulnerability 0.39, compared with 0.18
for tool-use nodes in the same structural group. This measured distinction
motivates conditioning error impact on both position and task type
when allocating model capacity and verification.

\noindent\textbf{Observation 2: Physical-plan decisions require joint
implementation and node-impact analysis.}
The value of an implementation depends on its combined model,
verification, and backend costs, its effect on the workflow's critical
path, and the consequence of errors at its node. These effects must
be evaluated under a common workflow-level objective.

\subsection{Physical Plans Must Remain Revisable}
\label{sec:motivation:lifecycle}

Even a coordinated physical plan selected at admission can become
unsuitable during execution. Offline profiles and target service rates
provide nominal service estimates; concurrent requests, workload shifts,
and backend congestion change the effective cost of the selected
implementations.

Our burst experiment illustrates this change
(\cref{sec:eval:repair}). The same balanced plan's burst p95 rises
from $135$ to $292$ seconds when the mix shifts toward long code
tasks. The balanced setting remains the best measured static setting:
its calibrated performance changes even though the best measured
preference does not. Tool failures can additionally require
request-specific repair that was not known at admission.

These events occur after some calls have been dispatched.
Adaptation must coordinate changes to the remaining model,
verification, and backend assignments without disrupting committed
work. It must also keep planning overhead low on the load-response
path.

\noindent\textbf{Observation 3: Physical plans require low-overhead runtime adaptation.}
A plan selected at admission must remain revisable as fleet and
execution conditions change. Updates must preserve dispatched
work and maintain coordinated choices for the uncommitted suffix
with low planning overhead.

Together, these observations motivate a layer that combines
per-request plan selection, coordinated workflow-level decisions,
and low-overhead revision during execution. \Cref{sec:layer} presents the plan abstraction and the contract
between the compiler and runtime.

\section{\sysname{}: The Workflow Physical-Planning Layer}
\label{sec:layer}

\begin{figure}[t]
  \centering
  \includegraphics[width=\linewidth]{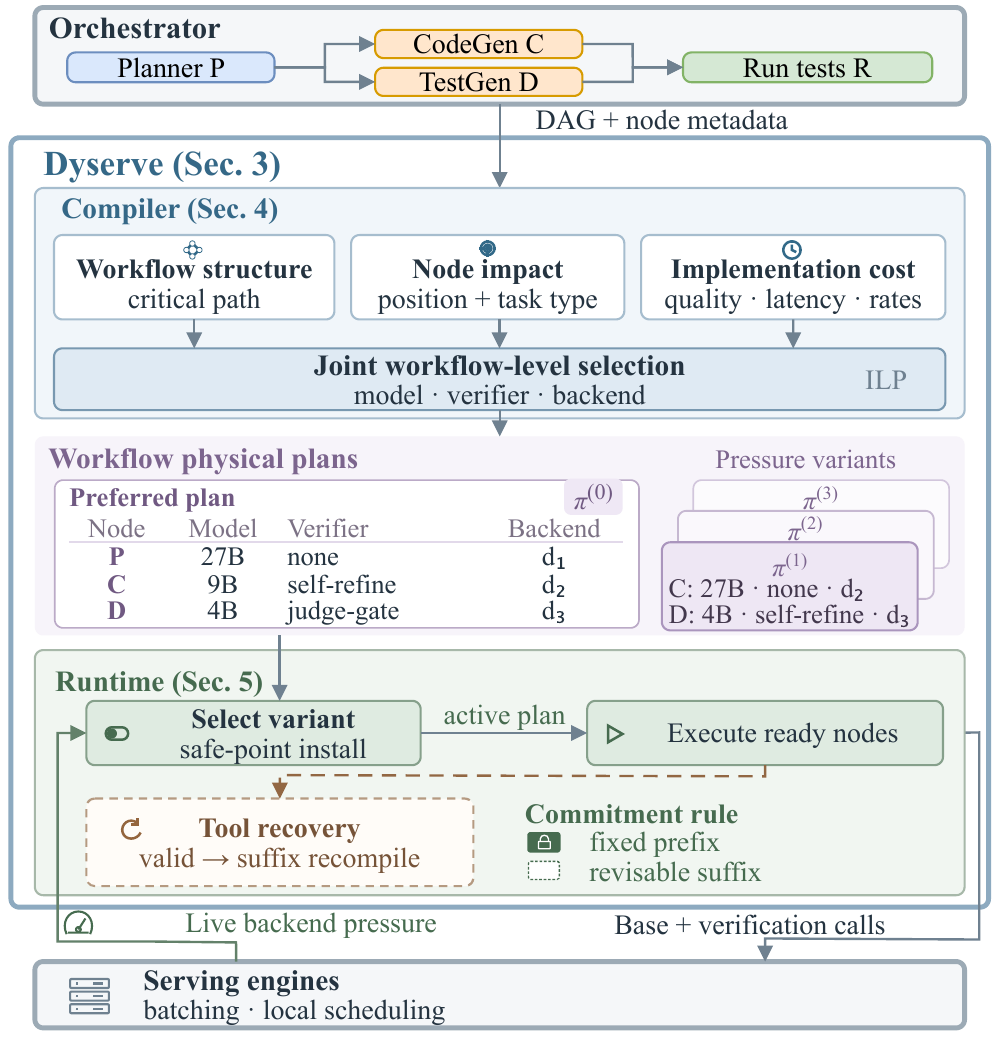}
  \caption{\sysname{} compiles and executes a materialized workflow.
  Profiles guide joint selection of a preferred physical plan and
  pressure-specialized variants. The runtime tracks workflow progress
  and uses backend feedback to adapt undispatched assignments.
  Bindings are illustrative.}
  \Description{The orchestrator submits a materialized workflow and
  node metadata to Dyserve. The compiler combines implementation
  profiles, position- and task-type-conditioned vulnerability, workflow
  structure, and calibrated service rates to construct physical plans.
  The runtime tracks dependencies and node completion, selects plan
  variants using live backend pressure, and dispatches base and
  verification calls. Dispatched bindings remain fixed. Successful
  bounded tool recovery can trigger one suffix recompilation.
  Serving engines return call results and pressure feedback.}
  \label{fig:compiler-runtime-contract}
\end{figure}

\sysname{} accepts a materialized workflow from the orchestrator and
manages its physical plan from admission through completion. Its
compiler jointly selects node implementations, while its runtime
executes the plan and adapts undispatched assignments to changing
serving load. \Cref{fig:compiler-runtime-contract} summarizes the flow
from workflow input to backend execution and load feedback.

\subsection{Workflow Input and Plan Abstraction}
\label{sec:layer:plan}

For each request, the orchestrator submits a finite materialized
\emph{directed acyclic graph} $G=(V,E)$ to \sysname{} for compilation
and execution. Nodes represent LLM operations or tool invocations,
and edges specify their dependencies. Branches are resolved and bounded
loops unrolled before submission. Per-node metadata identifies the
required skill and legal implementation candidates.

For an LLM node $n$, an execution option $o\in\mathcal{O}_n$ specifies
a model and serving replica (backend), while $p\in\mathcal{P}_n$ specifies a
verification policy, including no verification. Each pair $(o,p)$
defines a complete physical implementation. A \emph{workflow physical
plan} $\pi$ assigns an implementation to each LLM node and records the
application-defined tool invocations:
\begin{equation}
\begin{aligned}
    \pi(n) &\in \mathcal{O}_n \times \mathcal{P}_n,
        && n\in V_{\mathrm{LLM}},\\
    \pi(n) &= i_n^{\mathrm{tool}},
        && n\in V_{\mathrm{tool}}.
\end{aligned}
\label{eq:physical-plan}
\end{equation}
Here, $V_{\mathrm{LLM}}$ and $V_{\mathrm{tool}}$ partition $V$, and
$i_n^{\mathrm{tool}}$ denotes the tool invocation associated with node $n$.

\subsection{Compiler-Runtime Co-design}
\label{sec:layer:contract}

\noindent\textbf{Profile-guided joint compilation.}
To exploit the request-dependent headroom in Observation~1, \sysname{}
selects a plan for each request using reusable profiles rather than
executing candidate plans online. Skill-conditioned implementation
profiles estimate output quality and call demand. Vulnerability
profiles estimate error impact by node position and task type.
Calibrated backend service rates translate call demand into nominal
service costs, including the cost of verification on each candidate
model and backend. The compiler combines these estimates with workflow
precedence and downstream reach to account for the coupled costs and
node impact identified in Observation~2. A workflow-wide ILP then
jointly selects model-verifier-backend assignments, balancing
predicted quality and residual risk against critical-path latency and
serving cost (\cref{sec:admission}).

\noindent\textbf{Precompiled plan variants.}
Observation~3 requires adaptation without repeated optimization on the
load-change path. The initial solve produces a preferred plan
$\pi^{(0)}$. Additional solves at a bounded set of pressure-specialization
levels produce a plan bundle
$\mathcal{B}(G)=\{\pi^{(0)},\ldots,\pi^{(K-1)}\}$.
The variants provide alternative quality-demand trade-offs and carry
model-demand annotations for runtime selection. Prepared variants can
be cached for reuse across requests for the same workflow
(\cref{sec:admission:variants}). This separates workflow-wide plan
construction from low-overhead adaptation during execution.

\noindent\textbf{Plan execution and adaptation.}
The runtime maintains the active plan, node states, and intermediate
results for each request. It identifies ready nodes from the DAG's
dependencies and executes their assigned implementations. An LLM node
completes after its base call and selected verification policy finish;
the runtime records the resulting output and updates successor
readiness.

During execution, the runtime combines live backend pressure with the
compiled demand annotations to select a plan variant. At node-boundary
safe points, it applies that variant's assignments only to LLM nodes
whose base calls have not been dispatched. Dispatch commits the model,
verification policy, and backend for the entire invocation, including
verification stages that have not yet started. Completed and in-flight
bindings remain unchanged, and load-driven updates invoke no solver.
Tool-invocation errors follow a separate path: successful bounded
recovery can trigger one suffix recompilation with committed bindings
fixed (\cref{sec:runtime}).

\subsection{Serving-Engine Interface}
\label{sec:layer:serving}

The runtime submits base and verification calls to the backends
specified by the active physical plan. Serving engines execute these
calls and return outputs and completion status to the runtime.
Tool calls use their application-defined execution interfaces.

The serving fleet also exposes calibrated service rates for compilation
and live pressure signals for runtime adaptation. The rates support
nominal cost estimation; live pressure guides variant selection rather
than changing those estimates. Serving engines retain batching,
KV-cache management, and local scheduling. Once the workflow completes,
\sysname{} returns its final result to the orchestrator.

\section{Physical-Plan Compiler}
\label{sec:admission}

\begin{figure}[t]
  \centering
    \includegraphics[width=\linewidth]{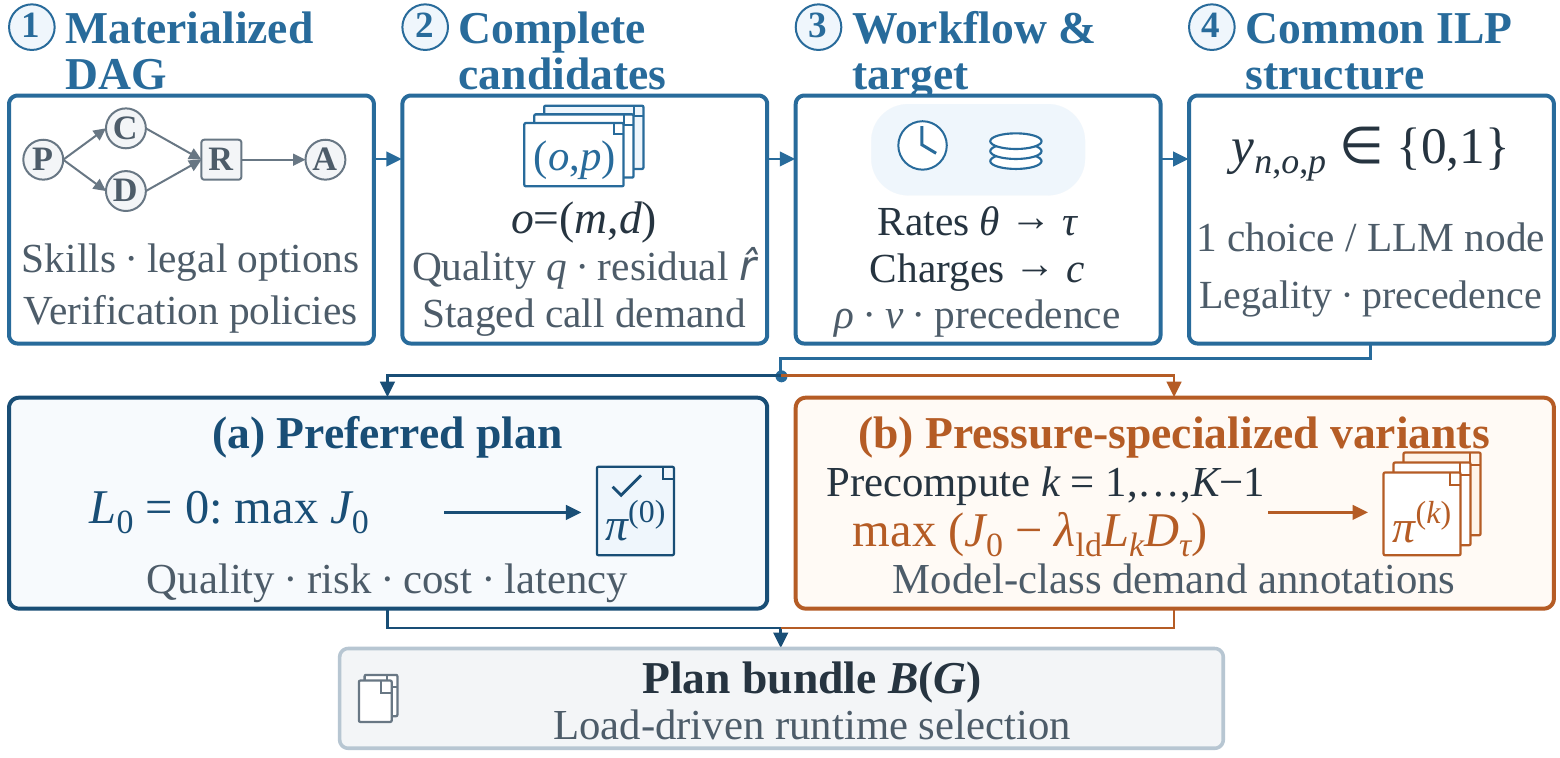}
    \caption{The compiler evaluates complete implementations using
    pair-specific quality and token-demand measurements, verification-call
    dependencies, workflow impact, and target service rates. The per-request
    ILP selects a plan; additional solves prepare pressure variants.}
  \label{fig:compiler-pipeline}
\end{figure}

For each arriving request, \sysname{} compiles the materialized
workflow into a physical plan under the contract of \cref{sec:layer}.
It binds measured model-verification profiles to candidate
implementations and converts their token demand into target-specific
service estimates. A workflow-wide ILP combines these estimates with
workflow structure and node vulnerability to select one implementation
for each LLM node. Additional solves prepare pressure-specialized
variants for runtime selection. \Cref{fig:compiler-pipeline}
summarizes this process.

\subsection{Pair-Specific Profiling and Candidate Binding}
\label{sec:admission:profiles}

To evaluate a workflow without executing its alternatives, \sysname{}
uses offline profiles indexed by skill class, model, and verification
policy. For each profiled skill class, we execute the candidate
model-verification pairs, including the no-verification configuration.
Each execution includes base generation and the calls required by the
selected policy. We record the resulting output quality, residual errors,
and per-call token demand, capturing the behavior of each complete pair.

At compilation, an LLM node $n$ supplies a dominant-skill tag $s_n$,
legal model-backend options $\mathcal{O}_n$, and verification policies
$\mathcal{P}_n$. For a candidate $o=(m,d)$ and $p$, the compiler retrieves
the corresponding $(s_n,m,p)$ measurements. Its quality coefficient is
$q_{n,o,p}=\widehat Q(s_n,m,p)$, where $\widehat Q$ denotes the measured
correctness rate of that pair's returned outputs. The pair profile also
estimates the error rate of the output returned by the complete
base-and-verification execution, denoted $\hat r_n(o,p)$. This is the
error rate remaining after the selected policy, including the base-only
case when no verification is chosen. Both estimates use the same
profiled samples and binary correctness criterion, so
$\hat r_n(o,p)=1-q_{n,o,p}$.
Nodes with the same skill tag reuse these measurements.
The token-demand profiles are converted into backend-specific service
estimates as described next. Probe datasets and the policy catalog are
specified in Appendices~\ref{app:verifiers} and~\ref{app:probes}.

\subsection{Verification Calls and Target Specialization}
\label{sec:admission:service}

The latency of a model-verification implementation includes its base
generation and all calls required by the selected verification policy.
For example, self-refinement generates an initial response, produces
feedback, and revises the response in sequence. A policy with multiple
critiques may execute them in parallel before a final revision.
The compiler uses these dependencies to estimate the duration of the
complete implementation rather than adding a fixed verification overhead.

The compiler represents each policy as an ordered set of stages
$\mathcal{S}_p$, including base generation. Each stage $a$ contains a
set $\mathcal{C}_a$ of calls that can execute in parallel; with no
verification, only the base-call stage remains. Let
$g^{a,j}(s_n,m,p)$ denote the profiled output-token count of call $j$
in stage $a$. For execution option $o=(m,d)$, the compiler converts
this demand into nominal service time using the calibrated decode
rate $\theta_{\mathrm{dec}}(m,d,b_0)$:
\begin{equation}
\tau_{n,o,p}
=
\sum_{a\in\mathcal{S}_p}
\max_{j\in\mathcal{C}_a}
\frac{g^{a,j}(s_n,m,p)}
     {\theta_{\mathrm{dec}}(m,d,b_0)}
+\tau_{\mathrm{pre}}.
\label{eq:tau}
\end{equation}
Here, $b_0$ is the calibration operating point, and
$\tau_{\mathrm{pre}}$ is the candidate's aggregate prefill-latency
correction on backend $d$. Serial stage durations add, while each
parallel stage contributes its longest nominal call duration.
Gated policies use expected demand based on measured gate rates.
The cost coefficient $c_{n,o,p}$ measures normalized token count across
all base and verification calls, including those that execute in parallel.

Separating pair-specific token demand from hardware-specific service
rates enables profile reuse across backends. With models, verification
policies, and inference settings unchanged, adapting to new hardware
requires only calibrating per-model decode throughput and prefill latency
on the target backend, rather than repeating the model-verification
profiling campaign. The compiler reuses the pair-level quality and
token-demand measurements with these updated service parameters.
Transient queueing is excluded from $\tau_{n,o,p}$; live backend pressure
guides runtime variant selection instead.

Equation~\eqref{eq:tau} accounts for serial and parallel calls within
one candidate node implementation. Section~\ref{sec:admission:selection}
then combines selected node durations through the workflow's
inter-node dependencies.

\subsection{Workflow-Aware Plan Selection}
\label{sec:admission:selection}

The preceding sections characterize each candidate implementation's
local quality, error rate, nominal duration, and call cost. Selecting a
workflow plan requires two additional assessments: the consequence of
an error at each node and the contribution of its execution time to the
workflow's critical path. The compiler combines these assessments before
jointly selecting the node implementations.

\noindent\textbf{Weighting local quality and error impact.}
The pair-specific error estimate $\hat r_n(o,p)$ describes the output
returned after base generation and the selected verification policy
have completed. It measures how often the complete implementation
returns an incorrect result, not how severely that result affects the
workflow. The vulnerability estimate $v_n$ supplies this second factor.
As defined in \cref{sec:motivation:coupling}, the compiler obtains it
from the node's structural role, task type, and fan-in:
\[
    v_n=\Psi\bigl(\operatorname{role}(n),s_n,
                  \operatorname{fanin}(n)\bigr),
    \qquad f_{n,o,p}=v_n\hat r_n(o,p).
\]
Appendix~\ref{app:vulnerability} gives the conditional estimator.
The resulting impact-weighted error penalty assigns greater value to
reducing errors at more vulnerable nodes. It uses measured pair-level
error rates and group-level fault-impact estimates as a planning
surrogate, rather than a calibrated probability of workflow failure.

The quality reward also uses a structural reach weight $\rho(n)$ to
emphasize nodes whose outputs feed more downstream operations. Let
$R(n)$ contain $n$ and all nodes reachable from it. For $|V|>1$, we use
$\rho(n)=1+\beta e(|R(n)|-1)/(|V|-1)$; for a single-node workflow,
$\rho(n)=1$. The effort parameter $e\in[0,1]$ adjusts this emphasis,
with $\beta$ bounding its magnitude. Reach is a structural weighting
rule, whereas vulnerability measures the consequence of injected faults.

\noindent\textbf{Accounting for parallel branches.}
A node's nominal duration already includes its base and verification
calls under \cref{eq:tau}. Workflow latency, however, is not the sum of
all node durations: serial dependencies accumulate time, while a join
waits for the last of its predecessors. For a fixed plan, let $d(n)$
denote the selected implementation's duration. The fork-join region
in \cref{sec:motivation:coupling} therefore contributes
$\max\{d(C),d(D)\}$, excluding the shared stages. Verification added to
the shorter branch increases this delay only when that branch exceeds
the other branch's completion time. The compiler evaluates this effect
through the longest path of nominal durations in the workflow DAG.

\noindent\textbf{Joint implementation selection.}
We encode the choice of a complete implementation with a binary
variable $y_{n,o,p}$, and set
$d(n)=\sum_{o,p}\tau_{n,o,p}y_{n,o,p}$ for LLM nodes.
Tool nodes retain their fixed duration estimates. Let $a_n\geq0$ be
auxiliary start times and $T\geq0$ the workflow completion-time
variable. The ILP imposes
\begin{equation}
\begin{aligned}
\sum_{o\in\mathcal{O}_n}\sum_{p\in\mathcal{P}_n}y_{n,o,p}
    &=1, && n\in V_{\mathrm{LLM}},\\
a_n &\geq a_u+d(u), && (u,n)\in E,\\
T &\geq a_n+d(n), && n\in V.
\end{aligned}
\label{eq:admission-constraints}
\end{equation}
The first constraint selects one implementation per LLM node. The
second applies to every incoming edge: at a join, it requires the
successor to wait for all predecessors, implicitly taking the maximum
of their completion times. The last bounds completion of the workflow.
With a positive latency penalty, the optimal $T$ for fixed assignments
is their nominal longest-path duration. These variables capture
precedence and potential overlap, not queueing or contention between
calls sharing a backend.

The preferred plan maximizes
\begin{equation}
\begin{aligned}
J_0 ={}&
\sum_{n\in V_{\mathrm{LLM}}}\sum_{o,p}
\bigl(\rho(n)q_{n,o,p}-\lambda_f f_{n,o,p}
-\lambda_c c_{n,o,p}\bigr)y_{n,o,p}
-\lambda_{\ell}T.
\end{aligned}
\label{eq:admission-objective}
\end{equation}
The objective rewards local quality, penalizes impact-weighted errors
and call costs, and charges workflow critical-path latency. Since
$\hat r_n(o,p)=1-q_{n,o,p}$, the quality and error terms equivalently
weight local correctness by $\rho(n)+\lambda_f v_n$, up to a
plan-independent constant. The same pair variable selects all of a
candidate's coefficients. A model or
verification change can therefore alter both local utility and the
critical path used to evaluate choices elsewhere in the workflow.
The configured weights express a quality-cost-latency preference,
not a hard end-to-end accuracy guarantee. They are selected on held-out
workflows and reused across the evaluation. Fixed tool quality and cost
terms are constants and are omitted; tool durations remain in the
precedence constraints. The selected assignments form $\pi^{(0)}$.

\subsection{Pressure-Specialized Plan Variants}
\label{sec:admission:variants}

Critical-path latency alone does not account for nominal execution
time hidden behind a longer branch. For example, verification that
fits within branch slack leaves $T$ unchanged but still adds calls.
To prepare alternatives with lower nominal demand, the compiler also
considers the sum of selected LLM-node durations,
$D_{\tau}=\sum_{n\in V_{\mathrm{LLM}}}d(n)$.
At each specialization level in the bounded set
$0=L_0<\cdots<L_{K-1}$, it maximizes
\begin{equation}
J_L=J_0-\lambda_{\mathrm{ld}}L D_{\tau}.
\label{eq:variant-objective}
\end{equation}
Unlike $T$, $D_{\tau}$ counts every LLM node, including nodes outside
the critical path. It is a nominal-duration demand proxy, not a direct
measurement of GPU utilization or the sum of all per-call service work:
parallel calls within a node have already been combined in
\cref{eq:tau}.

The measured pair profiles, calibrated service rates, and legality
constraints remain fixed across these solves. Increasing $L$ places
more weight on reducing nominal demand; it is not an instantaneous
queue measurement and does not require successive variants to use
different model classes. The runtime can skip variants that retain
the congested dominant model when selecting a less-loaded alternative
(\cref{sec:runtime:load}). Each variant carries a profile-based quality
estimate and per-model demand annotations identifying its dominant
model class. The runtime combines these annotations with live
class-level pressure, rejecting transitions whose estimated quality
drop from the active plan exceeds a configured tolerance. It updates
only undispatched assignments (\cref{sec:runtime:load}).
The preferred admission solve runs for every request. On a cache miss,
execution starts as soon as $\pi^{(0)}$ is ready, while additional variants
are prepared during execution. Load-driven switching becomes available
once those variants are ready. \Cref{sec:eval:overhead} reports the costs
of admission compilation and additional variant solves separately.
Algorithm~\ref{alg:gen} summarizes coefficient construction and
plan-bundle preparation.

\section{Adaptive Plan Runtime}
\label{sec:runtime}

The runtime executes the materialized workflow, tracks node completion,
and dispatches nodes when their dependencies are satisfied. It adapts
execution through two paths: backend load changes select a precompiled
plan variant, while tool errors trigger bounded recovery and one suffix
recompilation after a valid replacement is obtained. Requests share
backend-load signals but maintain separate physical plans and execution
states.

\subsection{Load-Driven Plan Adaptation}
\label{sec:runtime:load}

\noindent\textbf{Monitoring backend load.}
The runtime reads queue measurements from the serving backends and
summarizes the load of replicas serving the same model, such as 9B or
27B. Each plan variant identifies the model receiving the largest share
of its planned serving demand as its \emph{dominant model}
(\cref{sec:admission:variants}). The runtime uses the load of that model's
backends to decide whether to continue using the variant. This model
is identified during compilation, rather than selected from the current
queue measurements.

\begin{figure}[t]
  \centering
    \includegraphics[width=\linewidth]{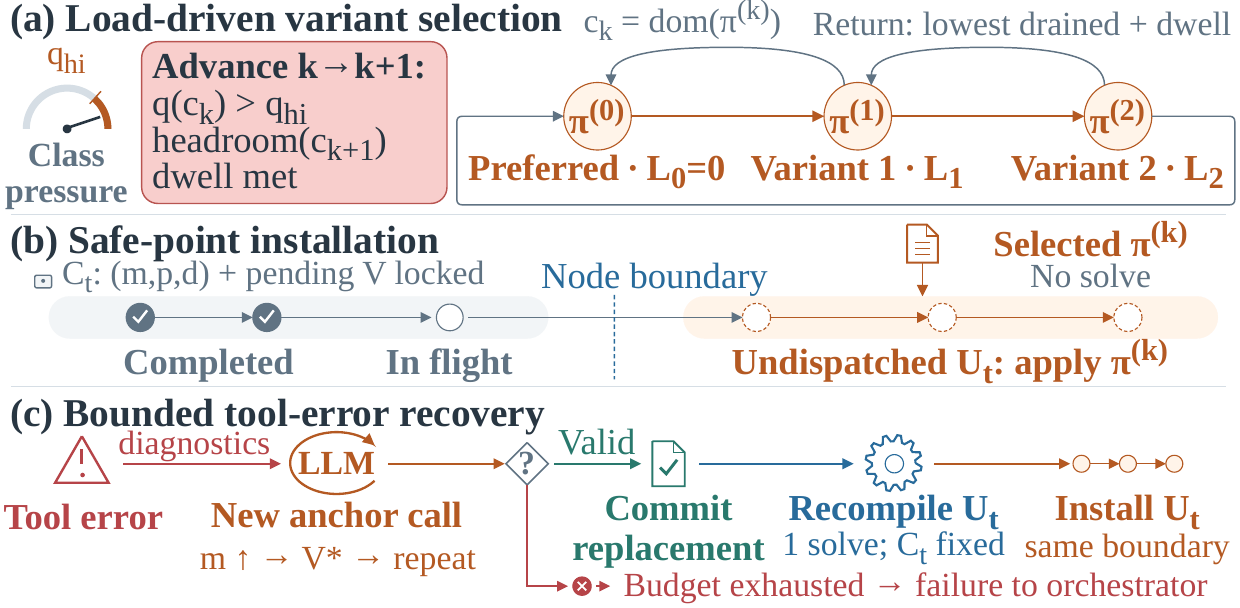}
    \caption{The runtime (a) selects precompiled variants under backend
    pressure, (b) updates undispatched bindings at node boundaries, and
    (c) recompiles the suffix once after successful bounded tool-error
    recovery. Dispatched bindings remain fixed.}
  \label{fig:runtime}
\end{figure}

\noindent\textbf{Switching plan variants.}
Variants are ordered by the pressure-specialization levels used during
compilation. When the load of the current variant's dominant model
exceeds the high-load threshold, the runtime scans later variants in
that order. It skips candidates whose dominant model is the same
congested class or is itself overloaded. Reducing verification while
retaining the congested dominant model is therefore insufficient.
A candidate must use a different dominant model with sufficiently
lower load under the configured comparison rule.

The runtime also compares the candidate's profile-based quality
(accuracy) estimate with that of the currently active plan. It rejects
a candidate whose estimated quality drop exceeds a configured tolerance,
even if the candidate's backend is idle. This limits the quality loss
of each transition. The runtime selects the first later variant that
passes both the load and quality checks, applying its precompiled
model-verifier-backend assignments together without invoking the optimizer.

When load subsides, the runtime returns directly to the lowest-index
variant whose dominant-model load is below the low-load threshold and
that passes the quality check.
A minimum dwell time between switches limits repeated changes in either
direction. If no transition is eligible, the current selection is retained.


\noindent\textbf{Applying the selected plan.}
Each request applies the selected variant at a node-boundary safe point,
updating the model, verification policy, and backend only for LLM nodes
whose base calls have not been dispatched. Once a base call is dispatched,
these choices remain fixed for the entire invocation, including
verification stages that have not yet started. Load-driven updates
neither re-execute completed nodes nor migrate or cancel in-flight calls
(\cref{sec:layer:contract}). The updated plan can therefore combine
previous assignments with the selected variant's assignments for the
remaining nodes (\cref{sec:walkthrough}).

\subsection{Bounded Recovery and Suffix Recompilation}
\label{sec:runtime:recovery}

Tool timeouts, malformed arguments, and explicit invocation errors
trigger recovery. The runtime identifies the anchor LLM node that
produced the failing invocation or its arguments and re-executes it
with diagnostic feedback under a bounded attempt budget. The recovery
sequence advances one model-size step, then selects the highest-quality
profiled verifier, and repeats the maximal configuration if the budget
permits. Each attempt is a separate invocation, not a change to a
previously dispatched node's assignment. Anchor operations must be safe
to re-execute. If the budget is exhausted, the runtime returns the failure
to the orchestrator.

After committing a valid replacement, the runtime invokes the compiler
once to select new implementations for the undispatched suffix. Completed
and in-flight assignments remain fixed. Completed outputs supply the
suffix's inputs, and the runtime waits for any in-flight predecessors
before dispatching their successors. Unlike load-driven switching,
this path solves the remaining planning problem rather than reusing a
precompiled variant. The residual formulation and its timing scope are
detailed in Appendix~\ref{app:residual}.

\section{A Physical-Plan Walkthrough}
\label{sec:walkthrough}

\begin{figure}[t]
  \centering
  \includegraphics[width=\linewidth]{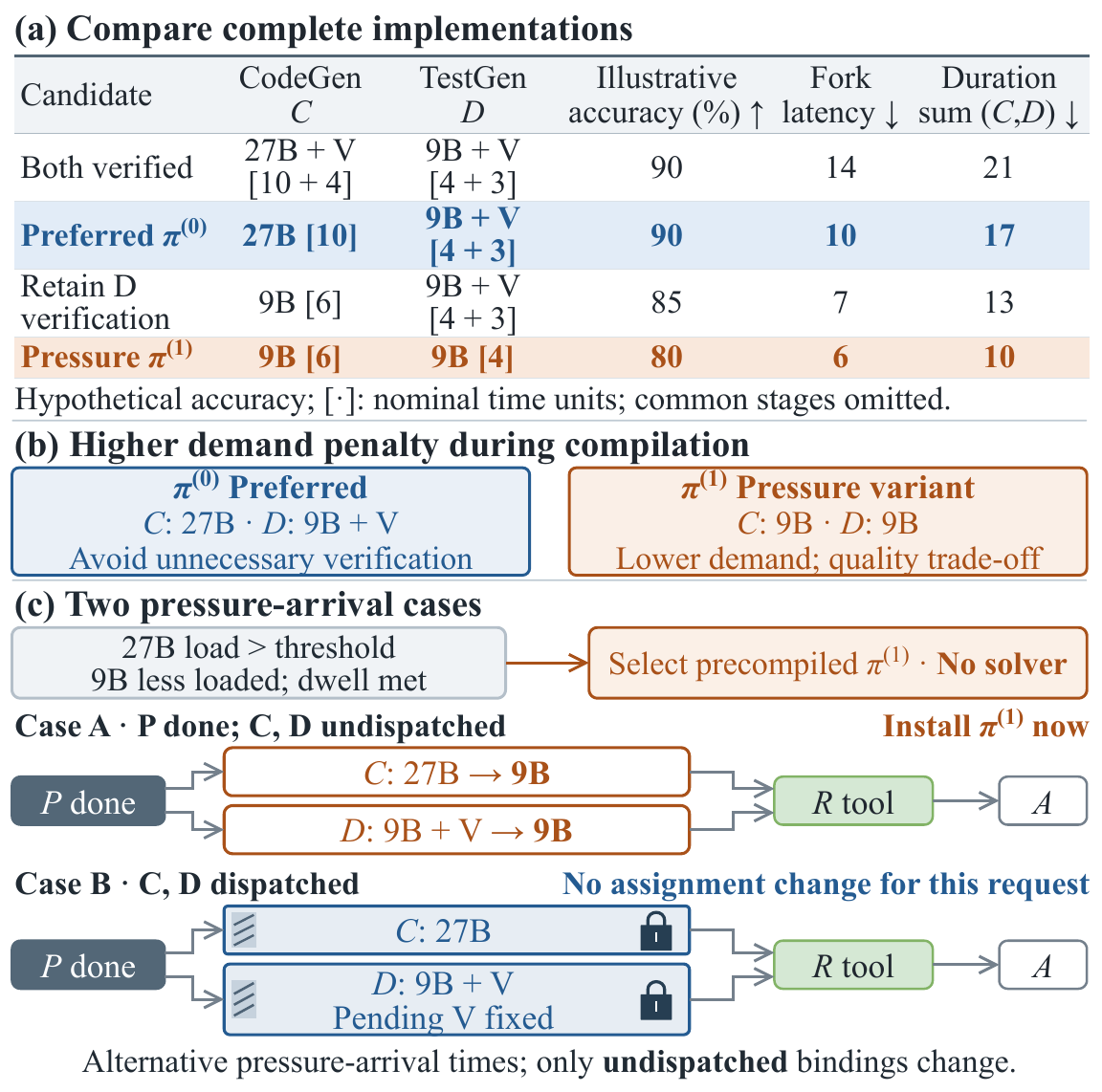}
  \caption{Joint plan selection and demand-aware adaptation. All values
  are illustrative; runtime updates change only undispatched assignments.}
  \label{fig:plan-walkthrough}
  \vspace{-5pt}
\end{figure}

Planner $P$ feeds parallel code generation $C$ and test generation $D$,
followed by test execution $R$ and answer node $A$.
\Cref{fig:plan-walkthrough}(a) uses hypothetical workflow accuracies and
nominal durations, not measurements or calibrated predictions.
$P$, $R$, and $A$ retain their assignments; their common costs and
durations are omitted.

\noindent\textbf{Joint plan selection.}
Suppose verification at $C$ provides no additional profiled quality or
risk benefit with 27B. The both-verified reference and $\pi^{(0)}$ have
90\% illustrative accuracy, but $\pi^{(0)}$ omits verification at $C$.
This reduces fork latency from 14 to 10 units and the branch-duration
sum from 21 to 17. Under the illustrated admission preference, the
compiler selects $\pi^{(0)}$ over the lower-quality 9B alternatives.

\noindent\textbf{Pressure-specialized selection.}
A higher specialization level increases the penalty on $D_\tau$, the
sum of nominal LLM-node durations, while retaining quality, risk, and
call-cost terms (\cref{sec:admission:variants}). In this example, that
penalty favors $\pi^{(1)}$: 9B without verification at both nodes,
with 80\% illustrative accuracy, six-unit fork latency, and a duration
sum of 10. Retaining verification at $D$ instead gives 85\%, seven
units, and a duration sum of 13. At the illustrated level, the additional
demand penalty outweighs its profiled benefit. The resulting variant
changes both $C$'s model and $D$'s verification policy.

\noindent\textbf{Load-driven update.}
The full-plan annotations identify 27B and 9B as the dominant models of
$\pi^{(0)}$ and $\pi^{(1)}$. A burst overloads 27B while 9B remains
less loaded. Suppose $\pi^{(1)}$ also passes the quality-drop check
relative to the active $\pi^{(0)}$. Once the load thresholds and dwell
conditions are met, the runtime selects $\pi^{(1)}$ without re-solving.
In Case~A, $P$ is complete and both branches are undispatched: the
runtime switches $C$ to 9B and removes $D$'s verification together.
In Case~B, both branches are dispatched, so their assignments and
pending verification remain unchanged. With $R$ and $A$ unchanged,
the selection has no effect on that request. Table values describe
complete plans, not partially updated executions.

\section{Evaluation}
\label{sec:eval}

We evaluate quality and latency, steady and bursty serving,
and recovery, followed by compiler ablations and control-path costs.
Admission-only experiments separate compilation from runtime adaptation.

\begin{figure*}[!t]
  \centering
  \includegraphics[width=\textwidth]{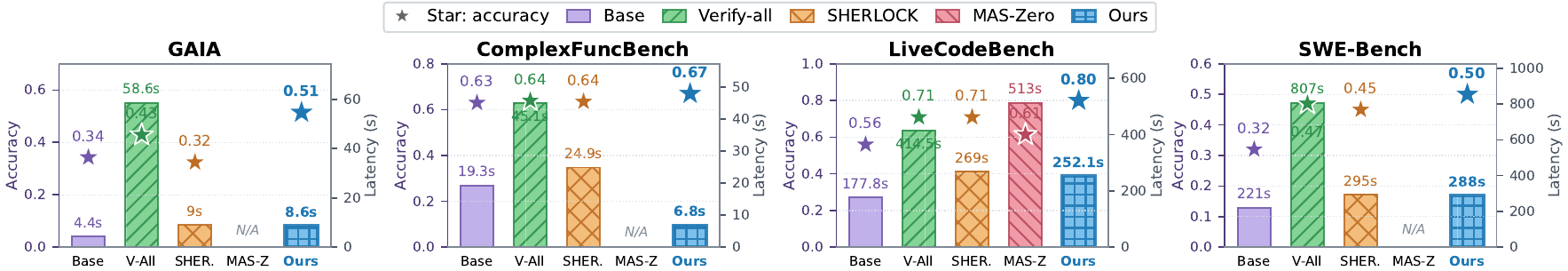}
  \caption{End-to-end accuracy (stars, left axis) and mean wall-clock
  latency (bars, right axis) on four workloads. SHERLOCK/SHER. denotes
  our training-free Sherlock-style approximation. MAS-Zero's unavailable
  cases are defined in \cref{sec:eval:setup}.}
  \Description{Four workload panels compare benchmark accuracy and mean latency for fixed-model execution, verification, Sherlock-style placement, MAS-Zero where supported, and Dyserve.}
  \label{fig:eval:end-to-end}
\end{figure*}

\subsection{Experimental Setup}
\label{sec:eval:setup}

\noindent\textbf{Workloads and platform.}
We evaluate LiveCodeBench (LCB)~\cite{jain2024livecodebench},
GAIA~\cite{mialon2024gaia}, ComplexFuncBench (CFB)~\cite{zhong2025complexfuncbench},
and SWE-bench~\cite{jimenez2024swebench} on 55, 35, 100, and 40 instances,
respectively, using their official correctness metrics. Timed-out or
incomplete requests count as incorrect. Flow~\cite{niu2025flow} with
GPT-5.4 generates workflows, whose materialized DAGs are frozen before
physical planning. The serving pool uses Qwen3.5-4B, Qwen3.5-9B, and
Qwen3.5-27B on NVIDIA H200 GPUs, with eight verification policies
(Appendix~\ref{app:verifiers}). Unless stated otherwise, quality and vulnerability
profiles use disjoint calibration data, with evaluated requests held out
from profiling. Service estimation follows \cref{sec:admission:service}.

\noindent\textbf{Baselines.}
\emph{Fixed-Model} (\emph{Base} in the single-request figure) uses 27B
without verification; \emph{Verify-All} adds the strongest profiled
verifier at every applicable node. Serving experiments use \emph{base
routing}: model-class-unaware round-robin call distribution.
\emph{Sherlock-style} uses a training-free approximation of Sherlock's
topology-aware placement and utility ordering~\cite{ro2025sherlock}.
\emph{MAS-Zero}~\cite{ke2026maszerodesigningmultiagentsystems} searches
logical workflows; other arms use the same materialized DAGs. All share
task inputs, endpoints, timeouts, and applicable verifiers. Our MAS-Zero
integration lacks GAIA/CFB tool support and exceeds the 4000-second
per-task search budget on SWE-bench; these cases are unavailable.

\noindent\textbf{Serving protocol and metrics.}
Four-GPU fleets replay mixed LCB, GAIA, and CFB traces with open-loop
Poisson arrivals and identical arrival sequences and deadlines across
arms. Balanced traces assign equal arrival mass to each workload.
SWE-bench is excluded because repository setup and tools dominate its
time. Fleet $(a+b)$ contains $a$ 27B and $b$ 9B replicas; 4B appears only
in single-request experiments. The primary and secondary fleets are
$(2+2)$ and $(3+1)$. Base routing and Sherlock-style default to $(4+0)$
unless labeled otherwise; each runtime comparison uses the fleet in
\cref{tab:eval:burst}.

End-to-end latency spans admission to final output, including queues,
model and verification calls, tools, and recovery. Model service time
is reported separately. Normalized SLO goodput is the fraction of arrivals
that are correct and finish by their deadline. Calm and burst windows
are assigned by arrival time. Main burst results use a 180-second target
selected from balanced-trace operating points. Each operating point
uses one trace unless noted; uncertainty methods and deadline sweeps
appear in Appendix~\ref{app:eval:additional}.

\subsection{End-to-End Quality and Latency}
\label{sec:eval:end-to-end}

Without load-driven switching, \sysname{} improves accuracy by
3-9 percentage points over each workload's highest-accuracy baseline
and achieves 1.1-6.8$\times$ speedups in mean latency
(\cref{fig:eval:end-to-end}).
On GAIA, accuracy increases from Verify-All's 0.43 to 0.51 while latency
falls from 58.6 to 8.6 seconds. On LCB, \sysname{} reaches 0.80 at
252.1 seconds versus Sherlock-style's 0.71 at 269 seconds.

\subsection{Serving under Steady Multi-Tenant Load}
\label{sec:eval:steady}

\begin{figure}[!htbp]
  \centering
  \includegraphics[width=\linewidth]{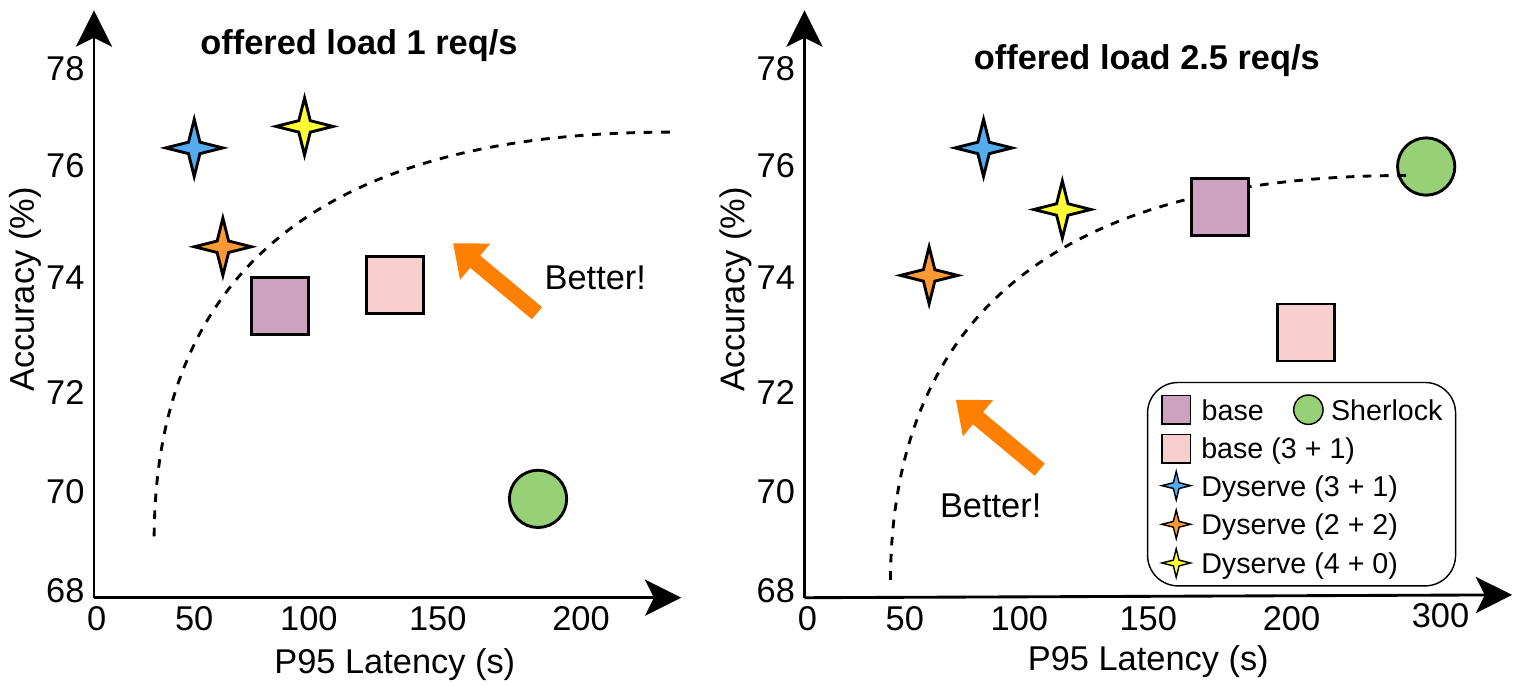}
  \caption{Steady serving at 1.0 and 2.5 requests/s: accuracy versus p95
  latency. Upper-left is better. Sherlock denotes our training-free
  Sherlock-style approximation; Base and Sherlock-style use $(4+0)$
  unless labeled otherwise.}
  \Description{At two offered loads, accuracy and p95 latency compare routing, verification placement, and compilation across four-GPU fleets.}
  \label{fig:eval:steady}
\end{figure}

\Cref{fig:eval:steady} isolates admission-time planning at 1.0 and
2.5 requests/s: plans remain fixed, and all arms use round-robin replica
routing. On $(4+0)$, compilation matches or improves base-routing
accuracy and reduces p95 from 172 to 112 seconds at 2.5 requests/s;
Sherlock-style reaches 291 seconds. Planning thus helps even without
smaller-model backends.

Heterogeneity requires appropriate allocation. Replacing one 27B replica
with 9B worsens base-routing p95 to 127/196 seconds at 1.0/2.5 requests/s,
with lower accuracy. Compiled plans on $(3+1)$ instead reach
56/86 seconds with higher accuracy. The plotted preference is
$\lambda_{\ell}=0.1$; sensitivity results appear in
Appendix~\ref{app:eval:sensitivity}.

\subsection{Runtime Adaptation under Bursts and Failures}
\label{sec:eval:repair}

\noindent\textbf{Burst protocol.}
We schedule pulses at 120-180 and 300-360 seconds, with a calm rate
of 0.3 requests/s and peaks of 4.0 on $(2+2)$ and 2.5 on $(3+1)$.
The primary trace's 400 arrivals end at 339 seconds, truncating the
second pulse. We compare base routing, a quality-leaning plan
($\lambda_{\ell}=0.05$) held fixed (\emph{admission only}) or with
switching, and the best measured static preference
($\lambda_{\ell}=0.1$) from the sweep. Balanced-trace variants use
$L\in\{0,1.5,3\}$. \Cref{tab:eval:burst} reports overall accuracy and
burst-window p95 and goodput.

\begin{table}[t]
\centering
\small
\setlength{\tabcolsep}{3pt}
\begin{tabular}{@{}llrrr@{}}
\toprule
Setting & Configuration & Acc. (\%) & p95 (s) & Goodput (\%) \\
\midrule
\multirow{3}{*}{\makecell[l]{Balanced\\mix, $(2+2)$}}
& Base routing & 70.8 & 336 & 25.7 \\
& Admission only & 75.8 & 823 & 18.1 \\
& With switching & 69.2 & 136 & 67.2 \\
\midrule
\multirow{3}{*}{\makecell[l]{Balanced\\mix, $(3+1)$}}
& Base routing & 72.5 & 201 & 62.9 \\
& Admission only & 76.2 & 229 & 71.0 \\
& With switching & 75.0 & 185 & 72.6 \\
\midrule
\multirow{2}{*}{\makecell[l]{Code-heavy\\mix, $(2+2)$}}
& Admission only & 66.0 & 599 & 35.7 \\
& With switching$^{\dagger}$ & 64.3 & 313 & 57.3 \\
\bottomrule
\end{tabular}
\caption{Runtime adaptation. Accuracy covers the full trace; p95 and
normalized goodput cover burst arrivals at the 180-second target.
All arms use the listed fleet. On $(3+1)$, differences between admission only
and switching reflect run variation because no switches occur.
$^{\dagger}$The code-heavy
configuration uses $L\in\{0,0.5,1\}$.}
\label{tab:eval:burst}
\vspace{-5pt}
\end{table}

\begin{figure}[!htbp]
  \centering
  \includegraphics[width=\linewidth]{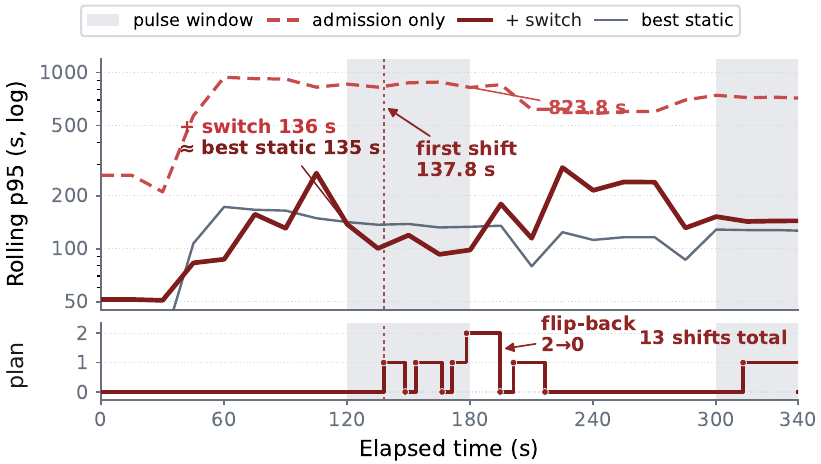}
  \caption{Burst adaptation on $(2+2)$. Top: rolling p95 in 60-second
  arrival windows (log scale). Bottom: active variant, with 13 switches
  and a return to the preferred plan ($0$). Admission only uses
  $\lambda_{\ell}=0.05$; best static uses $\lambda_{\ell}=0.1$.}
  \Description{The upper panel compares rolling p95 under a double-pulse burst. The lower panel records 13 plan switches and a return to the preferred plan after pressure drains.}
  \label{fig:eval:repair}
\end{figure}

\noindent\textbf{Adapting the quality-performance trade-off.}
On $(2+2)$, switching reduces burst p95 from 823 to 136 seconds and
raises goodput from 18.1\% to 67.2\%, versus 73.7\% for the best measured
static preference. During bursts, accuracy decreases from 75.4\% to
67.8\%, while deadline attainment increases from 23.1\% to 96.8\%.
On the representative trace, switching retains 77.6\% accuracy for calm
arrivals, compared with 72.4\% for the best static preference. It thus
adjusts the quality-performance trade-off with load rather than using
one lower-demand plan throughout (\cref{fig:eval:repair}). The burst-tail
benefit over base routing and admission only recurs across three arrival
seeds; the paired goodput test on the primary trace gives $p<10^{-4}$.

\noindent\textbf{Other fleets and workload mixes.}
No switches occur on $(3+1)$, so admission-only versus switching
differences reflect run variation. On the code-heavy trace, switching
raises goodput from 35.7\% to 57.3\% and reduces p95 from 599 to
313 seconds. This configuration uses different specialization levels,
$L\in\{0,0.5,1\}$, rather than a bundle frozen across mixes.

\noindent\textbf{Tool-failure recovery.}
We force the first invocation at 49 eligible tool sites in frozen GAIA
workflows to fail, with one trial per site and retry budget. One attempt
recovers 55\% of local failures; two recover 84\%, increasing mean trial
latency from 38.4 to 52.6 seconds. Terminal correctness rises from 14\%
without recovery to 20\% with either budget. This compares attempt
budgets, not the isolated benefits of escalation or suffix recompilation.

\subsection{Compiler Contributions and Model Validity}
\label{sec:eval:analysis}
\label{sec:eval:joint}

\Cref{tab:eval:ablation} combines joint-selection and compiler ablations
on 55 LCB inputs: 54 yield materialized workflows, and one shared
generation failure counts as incorrect for every method. The exhaustive
4B/27B group reports total model service time; the current-pool group
reports mean wall-clock latency. The joint result is shared by the
sequential and single-axis comparisons.

\begin{table}[t]
  \centering
  \small
  \begin{tabular}{@{}lrr@{}}
    \toprule
    Configuration & Accuracy (\%) & Time (s) \\
    \midrule
    \multicolumn{3}{@{}l}{\emph{4B/27B exhaustive study: model service time}} \\
    Model-first                    & 74.5 & 79.8 \\
    Verification-first             & 65.5 & 65.7 \\
    Model-only ILP                 & 60.0 & 35.4 \\
    Verifier-only ILP, 27B fixed    & 78.2 & 195.5 \\
    Verifier-only ILP, 4B fixed     & 52.7 & 12.2 \\
    Uniform topology weights       & 61.8 & 37.1 \\
    \textbf{\sysname{} (joint)}     & 76.4 & 37.2 \\
    \midrule
    \multicolumn{3}{@{}l}{\emph{Current pool: mean wall-clock latency}} \\
    No vulnerability term          & 70.9 & 247.6 \\
    Skill-collapsed profile         & 74.5 & 242.8 \\
    \textbf{\sysname{} (full)}      & 80.0 & 252.1 \\
    \bottomrule
  \end{tabular}
  \caption{Compiler comparisons on LCB. Both groups use the same
  55-input cohort but report different timing metrics, as labeled.}
  \label{tab:eval:ablation}
  \label{tab:eval:joint}
\end{table}

\noindent\textbf{Joint versus sequential selection.}
Joint and sequential methods share workflows, candidate pool, profiles,
and objective weights. Model-first selects models with verification
disabled, then selects verification with the models fixed.
Verification-first selects policies with all LLM nodes on 27B, then
optimizes models. Each stage is workflow-wide, but neither method
revisits its first-stage choices. Joint compilation provides
$2.15\times$ service-time speedup over model-first with 1.8 points higher
accuracy, and $1.77\times$ speedup over verification-first with
10.9 points higher accuracy. These results support
joint selection for latency-aware planning rather than fixing one
decision dimension before evaluating its final combination.

\noindent\textbf{Access to implementation choices.}
Model-only fixes verification to the best profiled policy per skill;
verifier-only fixes the model. These single-axis controls restrict the
candidate implementations, whereas both sequential methods select
models and verification policies.

\noindent\textbf{Workflow impact and profiles.}
Uniform topology weights reduce accuracy from 76.4\% to 61.8\% at nearly
unchanged service time. In the current pool, removing vulnerability or
collapsing skill profiles lowers accuracy from 80.0\% to 70.9\% or
74.5\%, respectively, at similar wall-clock latency. Workflow-impact
information and skill-specific profiles thus improve implementation
selection. Removing the full vulnerability term does not isolate
task-type conditioning.

\noindent\textbf{Service-model ranking.}
Across 55 frozen LCB workflows with a median of 16 executed plans each,
predicted duration and measured latency have a median per-workflow
Spearman correlation of 0.67 and 75\% pairwise ordering agreement.
These measurements assess candidate ranking, not absolute queueing-latency
prediction. Results by latency separation and preference sweeps appear
in Appendix~\ref{app:eval:sensitivity}.

\subsection{Control-Path Overhead and Compiler-Runtime Partition}
\label{sec:eval:overhead}

Across 92 frozen workflows with at most 11 nodes, initial compilation
takes 56 ms at p95, and each additional variant solve takes 15 ms
(\cref{tab:eval:overhead}). Initial compilation includes template
conversion, coefficient construction, and CBC solving; additional
variant preparation is measured separately.

\begin{table}[t]
\centering
\small
\setlength{\tabcolsep}{4pt}
\begin{tabular}{@{}lrr@{}}
\toprule
Operation & p50 & p95 \\
\midrule
Initial admission compilation & 27 ms & 56 ms \\
Additional variant solve & 12 ms & 15 ms \\
Coefficient construction & 0.2 ms & 0.2 ms \\
Profile load, once per process & \multicolumn{2}{c}{10 ms} \\
\bottomrule
\end{tabular}
\caption{Control-path costs over 92 frozen workflows. Initial
compilation includes conversion, coefficient construction, and solving;
additional variants repeat the solve. One-time profile loading has no
reported percentile split.}
\label{tab:eval:overhead}
\end{table}

\noindent\textbf{Compiler-runtime partition.}
In a separate latency-only control-path experiment, precomputed variants
achieve 78.3-second p95, versus 173.0 seconds for in-process solving and
130.3 seconds for isolated solving. This supports separating plan
construction from load-driven selection. The isolated configuration
also differs in installation timing, so the comparison evaluates
complete control-path designs rather than process isolation alone.
The full comparison appears in Appendix~\ref{app:eval:partition}.

\noindent\textbf{Scope.}
The evaluation covers four-GPU fleets on one node, one family of serving
models, and materialized DAGs, not larger fleets or dynamic expansion.
Plan quality remains subject to dominant-skill, service-time, and
residual-makespan approximations (\cref{sec:admission,sec:runtime:recovery}).

\section{Related Work}
\label{sec:related}

\noindent\textbf{Model routing and workflow configuration.}
FrugalGPT and RouteLLM optimize query-level cost-quality
trade-offs~\cite{chen2024frugalgpt,ong2025routellm}.
Aragog adapts model configurations at each workflow stage using live
load~\cite{dai2025aragog}.
Murakkab jointly optimizes workflow configurations, including
structured-executor parameters, model choices, and resource provisioning,
using workflow-configuration and hardware profiles. It combines
epoch-level reconfiguration with reactive autoscaling~\cite{chaudhry2025murakkab}.
In contrast, \sysname{} composes reusable skill profiles and
node-vulnerability estimates to jointly select model-verifier-backend
assignments for each request over existing backends. Its runtime
revises the executing request's undispatched assignments, including
verification policies, through precompiled variants under live pressure,
without a new solve or deployment reconfiguration.

\noindent\textbf{Verification and placement.}
Self-Refine improves outputs through feedback, while Archon searches
compositions of models and inference-time
techniques~\cite{madaan2023selfrefine,saad2024archon}.
Sherlock derives topology-guided placement from fault injection,
learns cost-aware verifier selection, and overlaps verification with
speculative execution~\cite{ro2025sherlock}.
\sysname{} conditions vulnerability on both position and task type,
co-selects models and verification policies, and prices their combined
execution on each candidate backend.

\noindent\textbf{Backend and workflow scheduling.}
vLLM and SGLang optimize inference execution~\cite{kwon2023efficient,zheng2024sglangefficientexecutionstructured};
Parrot, Agentix, and Orla provide application-aware scheduling and
resource control~\cite{lin2024parrotefficientservingllmbased,luo2026agentix,shahout2026orla}.
These mechanisms complement \sysname{}'s quality-sensitive physical
planning, which leaves engine-local execution policies unchanged.

\section{Conclusion}
\label{sec:conclusion}

\sysname{} is a workflow physical-planning layer combining per-request
compilation with runtime variant selection. Its model, verifier, and backend choices remain revisable before
dispatch; successful bounded recovery can trigger suffix recompilation. Across four workloads, it improves
accuracy by $3$-$9$ points at $1.1$-$6.8\times$ mean-latency speedups than the
highest-accuracy baseline. Under bursts, variant switching restores an
oversubscribed plan's goodput to within $6.5$ points of the best measured
static plan.

\begin{acks}
This work was supported in part by CoCoSys, one of seven centers in JUMP 2.0,
a Semiconductor Research Corporation (SRC) program sponsored by DARPA.
\end{acks}

\bibliographystyle{ACM-Reference-Format}
\bibliography{ref}

\appendix

\section*{Appendix}

\section{Verification Policy Call Structures}
\label{app:verifiers}

\Cref{tab:call_structure} summarizes the eight policy configurations,
including no verification. A node's legal set $\mathcal{P}_n$ determines
which configurations are applicable. The table lists only the
verification stages; the complete implementation includes base
generation as well. For example, separate base, feedback, and rewrite
calls form three serial stages, whereas \texttt{none} has only the base
stage. This convention avoids mixing additional calls with total calls.

The profile records per-call token demand and serial/parallel stage
membership. Serial stage durations add, a parallel stage contributes
its longest nominal call, and call cost sums all calls
(\cref{eq:tau}). Sample multiplicity, round limits, and gating behavior
belong to the profiled policy configuration, rather than being inferred
from its name. Gated policies use measured branch frequencies for
expected demand; this does not make their realized duration constant.

\begin{table}[t]
  \centering
  \small
  \begin{tabular}{@{}lp{.53\linewidth}@{}}
    \toprule
    Policy & Verification after base generation \\
    \midrule
    \texttt{none}              & No verification calls \\
    \texttt{self\_refine}      & Single-pass refinement \\
    \texttt{self\_refine\_iter}& Feedback $\to$ rewrite; repeat until stopping or the round limit \\
    \texttt{advanced\_refine}  & Critique stage $\to$ rewrite \\
    \texttt{self\_consistency} & Parallel samples $\to$ selection \\
    \texttt{judge\_gate}       & Pointwise verdict \\
    \texttt{gated\_refine}     & Gate $\to$ refinement if rejected \\
    \texttt{debate}            & Opponent/proponent rounds $\to$ selection \\
    \bottomrule
  \end{tabular}
  \caption{Verification-policy catalog. Base generation precedes the
  listed stages; arrows denote serial dependencies.}
  \label{tab:call_structure}
\end{table}

\section{Compilation and Plan-Bundle Construction}
\label{app:pseudocode}

Algorithm~\ref{alg:gen} makes the two profile inputs explicit:
$\Phi$ supplies model-verification measurements, and $\Psi$ supplies
position- and task-type-conditioned vulnerability. Each execution
option $o=(m,d)$ binds both a model and a backend. Coefficients and
legality constraints are constructed once for the bundle; only the
specialization level changes between its solves.

\begin{algorithm}[t]
  \caption{Construct a workflow physical-plan bundle.}
  \label{alg:gen}
  \begin{algorithmic}[1]
    \REQUIRE DAG $G=(V,E)$ and node metadata; profiles $\Phi,\Psi$;
      service rates $\theta$; objective weights and effort;
      levels $0=L_0<\cdots<L_{K-1}$
    \ENSURE Plans $\pi^{(0)},\ldots,\pi^{(K-1)}$ with demand and quality annotations
    \FOR{$n\in V_{\mathrm{LLM}}$}
      \STATE Compute $\rho(n)$ from reach in $G$ and request effort
      \STATE $v_n\gets\Psi(\operatorname{role}(n),s_n,
                            \operatorname{fanin}(n))$
      \FOR{$(o,p)\in\mathcal{O}_n\times\mathcal{P}_n$}
        \STATE Bind $(q,\hat r,\tau,c)_{n,o,p}$ using $\Phi$ and $\theta$
        \STATE $f_{n,o,p}\gets v_n\hat r_n(o,p)$
      \ENDFOR
    \ENDFOR
    \STATE Retain fixed tool invocations and duration estimates
    \STATE Build constraints \eqref{eq:admission-constraints}
    \FOR{$k=0,\ldots,K-1$}
      \STATE Solve \eqref{eq:variant-objective} at $L=L_k$
      \STATE Extract complete assignments $\pi^{(k)}$
      \STATE Attach per-model demand and estimated quality to $\pi^{(k)}$
    \ENDFOR
    \STATE \textbf{return} the indexed plan bundle
  \end{algorithmic}
\end{algorithm}

The ILPs use PuLP~\cite{mitchell2011pulp} with CBC. The $L_0=0$ solve
produces the preferred plan under \cref{eq:admission-objective}.
The algorithm describes bundle construction, not the timing of cache
population: on a cache miss, execution starts with the preferred plan
without waiting for additional variants (\cref{sec:admission:variants}).
Live pressure selects among prepared variants;
it does not rebuild these coefficients or invoke this procedure.

\section{Profiling Data and Vulnerability Estimation}
\label{app:probes}

\noindent\textbf{Skill probes and reuse.}
The skill profiles use held-out probes from HumanEval+ and
MBPP+~\cite{liu2023codegeneratedchatgptreally} for \textsf{code};
MATH-500~\cite{lightman2023lets}, AIME24, and
AIME25~\cite{dekoninck2026matharena} for \textsf{math}; and
GPQA-Diamond~\cite{rein2023gpqagraduatelevelgoogleproofqa} and
MMLU-Pro~\cite{wang2024mmluprorobustchallengingmultitask} for
\textsf{reasoning}. These measurements characterize complete
model-verification pairs. A new workflow reuses them through its nodes'
skill tags, provided those skill classes and configurations are covered.
The pair-quality probes and the workflow-level fault-injection trials
serve different purposes. Their calibration/evaluation separation
follows \cref{sec:eval:setup}.

\subsection{Conditional Vulnerability Estimator}
\label{app:vulnerability}

We follow Sherlock's counterfactual fault-injection
protocol~\cite{ro2025sherlock}: perturb a selected node, re-execute
downstream computations, and compare final correctness with the
unperturbed execution. We additionally condition the resulting impact
estimates on task type within each structural group.

Let $g=(\operatorname{role},s,\operatorname{fanin})$ identify a
profile group, and let $\mathcal{I}_g$ contain its paired unperturbed
and fault-injected trials. For trial $i$, let $Y_i^{(0)}$ and
$Y_i^{(1)}$ be binary final correctness without and with the injected
fault. The definition in \cref{sec:motivation:coupling} is
\begin{equation}
\begin{aligned}
  N_g^{+} &= \sum_{i\in\mathcal{I}_g}Y_i^{(0)},\\
  \Psi(g) &=
  \frac{\sum_{i\in\mathcal{I}_g}Y_i^{(0)}(1-Y_i^{(1)})}
       {N_g^{+}}, \qquad N_g^{+}>0.
\end{aligned}
\label{eq:vulnerability-estimator}
\end{equation}
Thus only baseline-correct pairs contribute to the denominator, and
$\Psi(g)\in[0,1]$ estimates how often an injected fault changes a
correct final result to an incorrect one. A group with $N_g^{+}=0$
has no empirical estimate; missing groups are not measurements of zero
vulnerability. Here, $N_g^{+}$ counts paired trials, not distinct workflows.

This conditional impact estimate is distinct from the natural error
rate of a model-verification pair. The compiler uses the latter from
$\Phi$, with $\hat r=1-q$ under the same pair-profiling samples and
binary correctness criterion,
and forms $f=v\hat r$ as a planning surrogate
(\cref{sec:admission:selection}). The injection protocol and reference
execution affect $v$; the product is not a calibrated workflow-failure
probability.

\section{Suffix Recompilation and Its Timing Scope}
\label{app:residual}

This procedure follows a successful bounded recovery, rather than a
load-threshold crossing (\cref{sec:runtime:recovery}). Let $C_t$ contain
completed and in-flight nodes whose bindings are committed, and let
$U_t=V\setminus C_t$ contain undispatched nodes. Dispatch commits the
entire LLM invocation, including verification stages that have not yet
started. The accepted recovery output supplies a fixed boundary input;
the recovery attempt is a separate invocation, not an overwrite of a
dispatched assignment.

The residual DAG is $G_t=G[U_t]$. Recompilation maximizes
\cref{eq:variant-objective} with its choice sums restricted to
$U_t\cap V_{\mathrm{LLM}}$. The selection and precedence constraints of
\cref{eq:admission-constraints} apply to the remaining nodes and edges;
tool invocations remain fixed. Let $\pi_t^{\mathrm{suffix}}$ denote
the resulting assignments on $U_t$. The updated plan is
\begin{equation}
  \pi_{t+1}
  = \left.\pi_t\right|_{C_t}\;\cup\;\pi_t^{\mathrm{suffix}}.
  \label{eq:suffix-repair}
\end{equation}
Here, $\operatorname{dom}(\pi_t^{\mathrm{suffix}})=U_t$.
The two assignment domains are disjoint; the merge rejects overlap
instead of replacing a committed binding. If $U_t$ is empty, there are
no remaining choices to optimize.

The residual makespan $T_t$ starts from zero on $G_t$ and spans only
undispatched work. Inputs from completed predecessors are available.
In-flight predecessors remain fixed boundary dependencies, and the
runtime waits for their actual completion before dispatching successors.
Their remaining time is not represented in $T_t$, so a long in-flight
branch can make it underestimate suffix completion time even before
queueing is considered. It is therefore a planning approximation, not
an end-to-end deadline guarantee.

Successful recovery triggers one residual solve. Load-driven switching
instead restricts an already compiled variant to $U_t$ and applies it
at a safe point, with no online solve (\cref{sec:runtime:load}).

\section{Additional Evaluation Results}
\label{app:eval:additional}

\subsection{Uncertainty and Additional Serving Results}
\label{app:eval:uncertainty}

\noindent\textbf{Uncertainty and replication.}
Each operating point uses one trace unless noted in \cref{sec:eval}.
We estimate within-trace uncertainty with request-level 95\% bootstrap
intervals and cross-check burst results with a 30-second block bootstrap
and paired sign-flip permutation tests on matched arrivals. On the
primary trace, the goodput intervals are 14.3-22.2\% for admission only
and 62.0-72.2\% with switching. Block resampling gives 9.5-25.7\% and
64.7-72.2\%, respectively; the paired test gives $p<10^{-4}$.
The burst-tail advantage over base routing and admission only recurs
across three independent arrival seeds, with at most 2.3 points of
per-arm overall-accuracy variation. On $(3+1)$, no switches occur and
paired tests yield $p\geq0.26$; admission-only versus switching
differences do not establish a switching benefit.

\noindent\textbf{Baseline comparisons.}
On LCB, MAS-Zero reaches 0.61 accuracy at 513 seconds, including
workflow search (\cref{fig:eval:end-to-end}). In steady serving, base
routing on $(4+0)$ has p95 latencies of 93 and 172 seconds at 1.0 and
2.5 requests/s. Replacing one 27B replica with 9B increases these to
127 and 196 seconds, while compiled plans on $(3+1)$ reach 56 and
86 seconds with higher accuracy. On the primary burst trace, the best
measured static preference has 135-second burst p95, versus base
routing's 336 seconds, with 4.7 points higher overall accuracy.

\noindent\textbf{Workload mix and deadline sensitivity.}
Under the code-heavy mix, the balanced preference remains the best
measured static setting, but its burst p95 rises from 135 to 292 seconds.
Using $L\in\{0,0.5,1\}$, switching reaches 57.3\% burst goodput, within
3.5 points of the best static result. These levels differ from the
balanced trace's $L\in\{0,1.5,3\}$, so this is not a frozen-bundle
comparison across mixes. We sweep deadlines from 60 to 300 seconds;
the main text uses a representative 180-second target selected from
balanced-trace operating points. Switching improves burst goodput over
admission only by 27-60 points on the primary trace and 11-33 points
on the code-heavy trace.

\noindent\textbf{Recovery by tool type.}
In the 49-site GAIA experiment, increasing the attempt budget from one
to two raises local recovery from 55\% to 84\%, while final correctness
remains 20\% with either budget. Additional recovery is concentrated
at web-fetch sites, where recovery rises from 14\% to 76\%.
The comparison varies attempt budgets; it does not isolate the effects
of escalation or suffix recompilation.

\subsection{Service-Model Ranking and Preference Sensitivity}
\label{app:eval:sensitivity}

\noindent\textbf{Candidate ordering.}
Across 55 frozen LCB workflows with a median of 16 executed plans each,
the per-workflow Spearman correlation between predicted duration and
measured latency has a median of 0.67. Pairwise ordering agreement is
75\% overall, increasing to 86\% for measured-latency separations of
at least $2\times$ and 90\% at $3\times$. The estimates distinguish
widely separated candidates more reliably than near-ties. These are
ranking measurements, not absolute queueing-latency predictions:
nominal durations exclude transient queues and output-length variation
(\cref{sec:admission:service}).

\noindent\textbf{Preference sensitivity.}
Sweeping effort from 0 to 1 moves accuracy from 0.71 into a
0.76-0.81 band while mean latency rises from 167 to 244 seconds.
In the steady heterogeneous-fleet experiment, $\lambda_{\ell}=0.05$
produces p95 latencies of 883-1179 seconds; $\lambda_{\ell}=0.1$
yields the operating point plotted in \cref{fig:eval:steady}.
Raising it to 0.2 reduces tail latency at a cost of 3-8 accuracy points.
These trade-offs motivate runtime adaptation when a quality-leaning
plan encounters higher load.

\subsection{Complete Control-Path Comparison}
\label{app:eval:partition}

The separate control-path experiment compares static execution,
in-process solving, isolated solving, and precomputed variants.
\Cref{tab:eval:partition:full} reports p95 end-to-end latency and
120-second deadline attainment. Attainment measures only completion
by the deadline, not correctness, and therefore differs from normalized
SLO goodput.

\begin{table}[t]
  \centering
  \small
  \begin{tabular}{@{}lrr@{}}
    \toprule
    Configuration & p95 (s) & Attainment (\%) \\
    \midrule
    Static execution       & 135.0 & 92.7 \\
    In-process solving     & 173.0 & 86.0 \\
    Isolated solving       & 130.3 & 92.0 \\
    Precomputed variants   & 78.3  & 98.3 \\
    \bottomrule
  \end{tabular}
  \caption{Control-path comparison at a 120-second deadline. Deadline
  attainment does not include correctness.}
  \label{tab:eval:partition:full}
\end{table}

The results support separating plan construction from load-driven
selection in the evaluated system. The isolated configuration also
differs in installation timing, so this is a comparison of complete
control-path designs rather than process isolation alone.

The compiler-overhead measurements cover 92 frozen workflows with
at most 11 nodes. Initial compilation includes template conversion,
coefficient construction, and CBC solving, but excludes additional
variant preparation. Load-driven switching invokes no solver;
successful recovery can recompile the undispatched suffix. No separate
switch latency or universal residual-solve bound is reported.

\end{document}